\documentclass[11pt]{article}

\usepackage[utf8]{inputenc}
\usepackage{centernot}
\usepackage{mathrsfs}
\usepackage{amsthm}
\usepackage{pifont}
\usepackage{marvosym}
\usepackage{geometry} 
\usepackage{amsmath}
\usepackage{graphicx} 
\usepackage[parfill]{parskip} 
\usepackage{booktabs} 
\usepackage{array}
\usepackage{paralist} 
\usepackage{verbatim} 
\usepackage{subfig} 
\usepackage{amssymb}

\usepackage{fancyhdr}
\usepackage{sectsty}
\usepackage[nottoc,notlof,notlot]{tocbibind} 
\usepackage[titles,subfigure]{tocloft} 
\title{A Stronger Foundation for Computer Science and P=NP}

\author{Mark Inman, Ph.D. \\ mark.inman@egs.edu}
\date{April 21, 2018}

\begin{document}
\maketitle
\abstract{This article describes a Turing machine which can solve for $\beta^{'}$ which is RE-complete. RE-complete problems are proven to be undecidable by Turing's accepted proof on the Entscheidungsproblem. Thus, constructing a machine which decides over $\beta^{'}$ implies inconsistency in ZFC. We then discover that unrestricted use of the axiom of substitution can lead to hidden assumptions in a certain class of proofs by contradiction. These hidden assumptions create an implied axiom of incompleteness for ZFC. Later, we offer a restriction on the axiom of substitution by introducing a new axiom which prevents impredicative tautologies from producing theorems. Our discovery in regards to these foundational arguments, disproves the SPACE hierarchy theorem which allows us to solve the P vs NP problem using a TIME-SPACE equivalence oracle.}

\section{A Counterexample to the Undecidability of the Halting Problem}
\subsection{Overview}
\subsubsection{Context}

Turing's monumental 1936 paper ``\emph{On Computable Numbers, with an Application to the Entscheidungsproblem}'' defined the mechanistic description of computation which directly lead to the development of programmable computers. His motivation was the logic problem known as the Entscheidungsproblem, which asks if there exists an algorithm which can determine if any input of first order logic is valid or invalid. After defining automated computing, he posited the possibility of a program called an $\mathscr{H}$ Machine which can validate or invalidate its inputs based on reading a description number for some given program description. However, while he constructed a Universal Turing Machine, he did not provide an actual construction of his $\mathscr{H}$ machine, only posited that one should exist. When he asked if this machine could decide for any input, he was able to show that in fact, it couldn't. His proof specifically depends upon this $\mathscr{H}$ Machine not being able to validate itself. He gives a detailed description as to why it can not, explained later in this article.

However, a close reading of his paper shows an added assumption by Turing when he constructs his $\mathscr{H}$ machine. While this assumption does not affect the construction or effectiveness of a Universal Turing Machine, it does have an affect on the overall results regarding the Halting problem and its sister problem, the Entscheidungsproblem, as well as any related results having to do with computability.

In this article, we give a detailed description on how to construct a self-validating $\mathscr{H}$ machine. The construction of a self-validating $\mathscr{H}$ machine may have application in fault-tolerance of run-time self-correcting code validation in artificial intelligence implementations. It may also lead to a better understanding of complexity relationships between complexity classes. It also expands our understanding of the theoretical limits of computation. 

\subsubsection{Preliminary Considerations}

The terms $\emph{Circular Machine}$ and $\emph{Circle-free Machine}$ are suitable for our description and we will use Turing's own definition of a computing machine. A $\emph{Circular Machine}$ is deemed unsatisfactory due to forever looping, redundantly over a repeating pattern. Also, a $\emph{Circle-free Machine}$ is satisfactory because of its ability to continue deciding indefinitely, without entering an unbounded repeating loop. Turing's description of the Halting problem is completely mechanical, while many modern descriptions rely on an oracle, reduction to Cantor's Diagonalization or logical reduction similar to G\"odel's Diagonalization Lemma. Using his terminology helps the reader directly compare this article with the original proof without intermediary interpretations or simplifications. However, conventionally,  a Circle-free Machine is considered to \emph{halt}, while a Circular Machine \emph{does not halt}. \cite{Turing}

A $\emph{Standard Description}$ or S.D. is the rule set for any given Turing Machine $\mathscr{M}$ in a standard form. By creating a standard, the rule sets themselves can be used to create a $\emph{Description Number}$ or D.N. which itself may be readable by a Universal Turing Machine, $\mathscr{U}$, as an instruction set. \cite{Turing} \\

\noindent{\emph{Remark.} A Description Number arbitrarily represents a Standard Description. Thus, we can choose which D.N. is used to represent some specific S.D. to our liking, as long as our constructed machine can read the D.N. and interpret it as the corresponding S.D. Again, D.Ns. are arbitrary, thus, if the D.Ns. we receive do not fit our format, where the ordering is consequential to the functioning of $\mathscr{H}_s$, when developing $\mathscr{H}_s$, we can re-assign new D.Ns. (which are not fixed\footnote{The description numbers are only fixed relative the construction of $\mathscr{H}_s$; we can always re-configure $\mathscr{H}_s$ to change any given D.N. however we wish, as such we regard any D.N. as not fixed.}) to the respective S.Ds. (which are fixed) such that $\mathscr{H}_s$ reads the given D.Ns. in the proper order in relationship to c (c, a natural number of relative size defined in section 1.2.1).}

From Turing's paper:  ``Let $\mathscr{D}$ be the Turing Machine which when supplied with the Standard Description (S.D.) of any computing machine $\mathscr{M}$ will test this S.D. and if $\mathscr{M}$ is circular will mark the S.D. with the symbol `$\emph{u}$' and if it is circle free, will mark it with `$\emph{s}$' for `unsatisfactory' and `satisfactory' respectively. By combining machines $\mathscr{D}$ and $\mathscr{U}$, we could construct a machine $\mathscr{H}$ to compute the sequence of $\beta^{'}$" \cite{Turing}

\subsubsection{Turing's Claim}
Turing claims that while $\mathscr{H}$ is circle free by construction, when $\mathscr{H}$ is given the description number for $\mathscr{H},$ it becomes circular. \cite{Turing} In the eighth section of Turing's paper on the Entscheidungsproblem, Turing claims that $\beta^{'}$ can not be determined because of the following reason:

``The instructions for calculating the R(K)-th [figure] would amount to `calculate the first R(K)-th figures computed by $\mathscr{H}$ and write down the R(K)-th'. This R(K)-th would never be found. I.e. $\mathscr{H}$ is circular..." \cite{Turing}

This is because, since $\mathscr{H}$ relies on certain subroutines to make its determination, when it reaches and tries to evaluate K, it must call itself, which provides instructions on reading inputs from 1 to K-1 in order to call the R(K)-th figure, but it can never get there, because it keeps repeating its own instruction loop. \cite{Turing} 

\subsubsection{Turing's False Assumption}
Turing assumed that his interpretation of $\mathscr{H}$, the one described just above, applies to all possible constructions of $\mathscr{H}$. Note that while he did formally define a Universal Turing machine, nowhere in Turing's paper did he actually formally define the full construction of $\mathscr{H}$. He assumed that any program with the property to determine the halting problem for a given input, would also not have any property which could learn when it enters a repeating loop for any given S.D. However, this is not necessarily the case and if we can provide an example of a program which does recognize a repeating loop, arbitrarily, such that it can switch states and act accordingly, then we've discovered a means to write $\mathscr{H}$ machine in such a way that it may solve for $\beta^{'}$. It only takes one positive example to universalize the example to all Universal Turing Machines. A negative example, such as the $\mathscr{H}$ machine assumed by Turing, is trivial upon the discovery of the existence of a positive example.

The problem reduces to describing a Turing Machine which can self-discover it's running its own instructions arbitrarily\footnote{by self-discovering its own instructions arbitrarily, we mean that it can recognize it is evaluating its own S.D. from the given D.N. Additionally, we are not referring to the existence of an initializer that feeds a fixed K to be recognized by a single read instruction that skips K and just ``rubber stamps'' approval. Such ``rubber stamping'' is considered a trivial case and is not of any meaningful concern.} when it reaches its Description Number (D.N.), such that some $\mathscr{H}$ machine configuration prints $\beta^{'}$.\\
\\
\noindent{\emph{Remark.} Because the D.N. is arbitrary for any S.D., there is no restriction on which specific D.N. can be used to represent some S.D. Thus, we may choose whichever D.N. pleases us to represent any given S.D., provided $\mathscr{H}$ can interpret the D.N. into the proper instructions.\cite{Lew} While our proof depends on the ordering of D.Ns., this is not a problem for our results because it only takes one ordering of D.Ns., which are arbitrary representations of all S.Ds, and this ordering can be universalized to any order by reordering after solving. Solving for $\beta^{'}$ means solving the halting problem on arbitrary input for any given S.D. There is no restriction on the order of verification of any S.D.

We will, in the next subsection, construct a \emph{Supermachine} that can recognize itself as its own input, from the S.D. simulation, which is then instructed to change to a circle-free state upon this recognition. Because such a construction exists, and because such a construction is arbitrary for any S.D. of this class of Turing Machines, we may solve for $\beta^{'}$ non-trivially.

\subsection{The Existence of Self-validating Computers}
\subsubsection{Supermachine} 

Let us consider that $\mathscr{H}^{'}$ is a controller machine with a D.N. of $K^{'}$. It controls two different $\mathscr{H}$ machines: $\mathscr{H}_0$ and $\mathscr{H}_1$. $\mathscr{H}_0$ and $\mathscr{H}_1$ each have the  ability to determine ``$\emph{u}$" or ``$\emph{s}$" on a D.N. input, except $\mathscr{H}_0$ tests as Turing describes, from D.N. 1 counting upwards (Each D.N. is a natural number) and $\mathscr{H}_1$ tests from a certain twos complement of whatever number is being tested by $\mathscr{H}_0$ as a simultaneous parallel input, such that its subsequent D.N. is one less than the previously tested D.N. Let us represent each D.N. by some integer $i$. $\mathscr{H}_0$ and $\mathscr{H}_1$ have a unique D.N. of $K_0$ or $K_1$ respectively.\footnote{This can be determined through a unique identifier string, which does not affect the machine's function or performance, but differentiates the two machines from each other giving them each a unique D.N.}

Upon input of any $i_0$ to be read by $\mathscr{H}_0$, let $\mathscr{H}^{'}$ store the value pair $(i_0, z)$ until $i_0$ is determined to be satisfactory or unsatisfactory. When the output is determined, let $\mathscr{H}^{'}$ replace the $(i_0, z)$ with the respective $(i_0, s)$ or $(i_0, u)$ in the data store, such that there is no longer a data store of  $(i_0, z)$. Let the same process occur for any $i_1$, such that $\mathscr{H}^{'}$ also initially stores each D.N. input with $(i_1, z)$ and $\mathscr{H}_1$ reads $i_1$ to determine satisfactory or unsatisfactory, subsequently replacing the initial value pair with the respective value pair $(i_1, s)$ or $(i_1, u)$ depending on the output of $\mathscr{H}_1$. A \emph{redundancy} occurs when some $i_0 = i_1$. 

Let $\mathscr{H}^{'}$ have the ability to compare value pairs such that the machine may recognize a redundancy when it occurs, and may also recognize when a value pair contains a z value on the condition of such a redundancy. Let's call this a \emph{z-check} ability.

Let $\mathscr{H}_s$ be the supermachine that is the configuration of all three $\mathscr{H}$ Machines as described above and let $K_s$ be the D.N. for the supermachine.

Initialize the identifier strings such that $K_1 < K_0$. %This is arbitrary, just need to have one K have a description number that is less than the other, i.e. they can't have the exact same configurations, even though they have the same decision abilities when run by a Universal Turing Machine, and let's make sure that we call that particular D.N., K_1.

Let the number of bits in $K_0 = n$. Let the twos complement of the first D.N. input to $\mathscr{H}_0$ , which is 1, be determined by $n$ such that it satisfies the equation $c = 2^n - 1$.
\\

\noindent{\emph{Lemma}. $\mathscr{H}_s$ proceeds circle free, until it reads $K_s$.} \\
If $c - K_0 > K_1$, then re-initialize the D.N.\footnote{one may re-initialize, if necessary, the D.N. by adding irrelevant description information into some S.D. yielding a different D.N. provided such information does not affect the integrity of the original S.D.} 
for either $\mathscr{H}_0$ or $\mathscr{H}_1$ such that $c - K_0 < K_1$. This guarantees that $\mathscr{H}_0$ will read $K_1$ before $\mathscr{H}_1$  reads $K_1$ and also guarantees $\mathscr{H}_1$  will read $K_0$ before $\mathscr{H}_0$ reads $K_0$. Let the controller $\mathscr{H}^{'}$ contain a memory command which stores the decision value pairs given by $\mathscr{H}_0$ and $\mathscr{H}_1$. The controller may routinely check for a redundancy on the next input. 

Now consider when $\mathscr{H}_0$ reads $K_1$, and $K_1$ calls the D.N. for $\mathscr{H}_0$: $K_0$ will call $K_1$, which will again call $K_0$ which will result in a z-check, recognizing that the value pair ($K_0$, z) is already stored in memory, and therefore, since $K_0 < c$, we know that $K_0$ is the description number for itself, is impossible to call by construction without calling $K_1$ first, which means it must be checking the description number for a machine which calls itself, namely $\mathscr{H}_1$, which allows us to correctly store the value pair ($K_1$, s). This same reasoning can be applied for when  $\mathscr{H}_1$ reads $K_0$, correctly storing the value pair ($K_0$, s). 

If however, the machine has determined a redundancy occurred on a value pair where the value is either $(i,s)$ or $(i,u)$ (i.e., a negative evaluation on the z-check, but the redundancy check is positive), then we have already evaluated this D.N. from the other $\mathscr{H}$ machine at the top level, and we no longer have to continue within the range 1 to $c$, since they will all have been decided. The supermachine, at this point proceeds to utilize machine $\mathscr{H}_0$ and proceeds from D.N. input value $c + 1$, and continues through the rest of all Description Numbers,  $c + 2$,  $c + 3$, etc... at least until it reaches its own D.N.,  $K_s$, for no other D.N. should be problematic\footnote{We should note here the significant finding by Yedidia and Aaronson of the independence of calculating BB(7918) from ZFC which will only halt if and only if ZFC is inconsistent. In a later section of this paper, we prove ZFC is in fact, inconsistent, meaning that BB(7918) is expected to eventually halt. We could thus expect $\mathscr{H}_s$ to determine that BB(7918) will halt. The unsolvability of the halting problem, as it is related to BB(7918) is contingent on ZFC being consistent, for BB(7918) will not halt if and only if ZFC is consistent. Forming such a Turing machine, which will halt if and only if the axiom set is inconsistent, using the proposed axiom in the later section of this article, is just not possible as the impredicative form of the machine will violate the proposed axiom.} in determining the output decision. Thus, $\mathscr{H}_s$ proceeds circle free, at least until it reaches $K_s$ which is easily constructed to be larger than $c$. $\square$\\

\begin{figure}\label{F:HMachine}
\includegraphics[width=\textwidth]{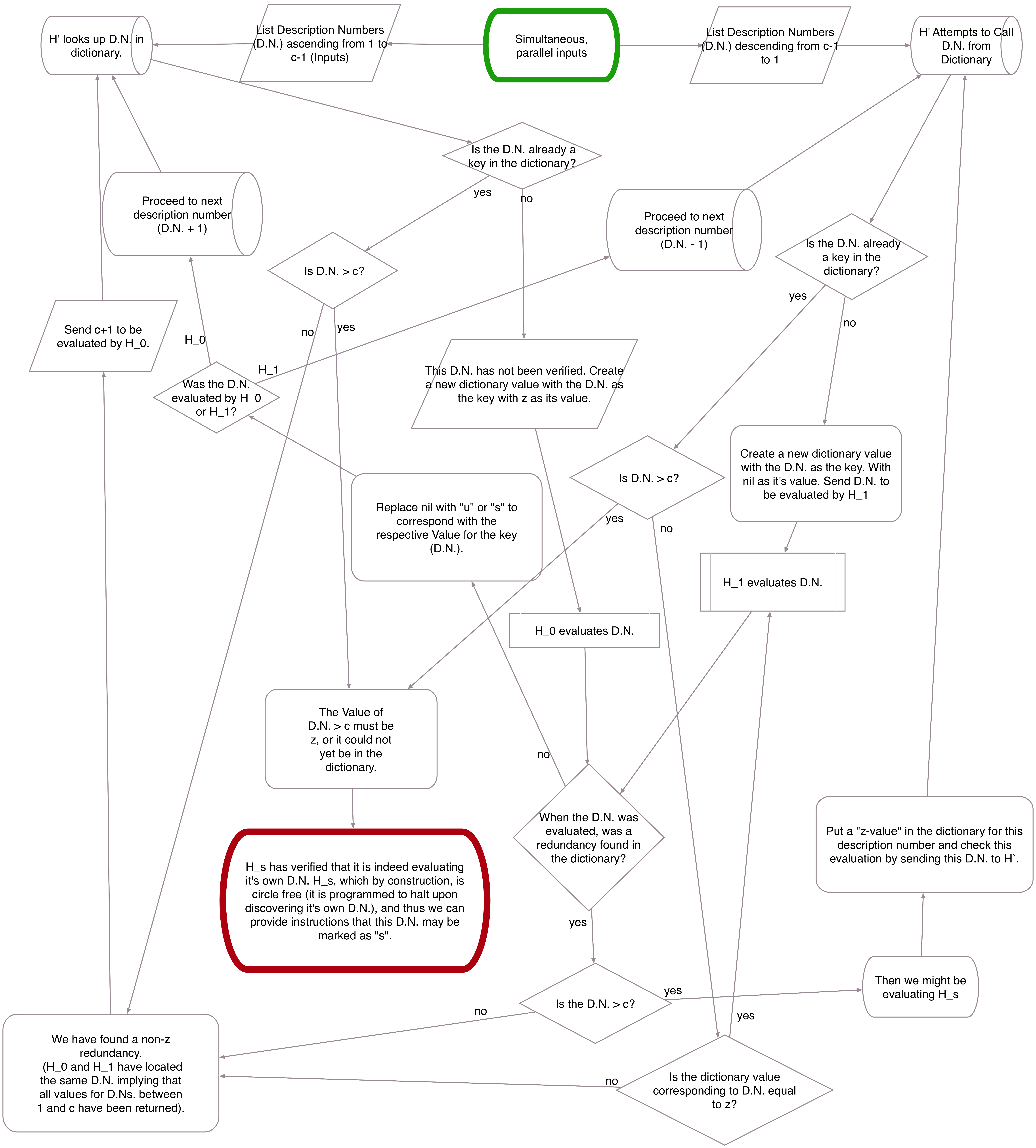}
\caption{A supermachine configuration appears to exist}
\end{figure}

\subsubsection{$\beta^{'}$ is Decidable}
\begin{proof}

\noindent\emph{$\beta^{'}$ is Decidable.} At the point $K^{'}$ is received as an input, it is determined satisfactory by either $\mathscr{H}_0$ or $\mathscr{H}_1$. Neither $K_0$ nor $K_1$ are called during this phase of the process.
 
By lemma, $K_0$ is decided by $\mathscr{H}_1$, $K_1$ is decided by $\mathscr{H}_0$ and $\mathscr{H}_s$  continues indefinitely until we reach $K_s$, which describes $\mathscr{H}_s$. $K_s$ is read by $\mathscr{H}^{'}$ and as before, its Description Number is stored along with its temporary pair value of z until $\mathscr{H}_0$ or $\mathscr{H}_1$ returns a value for $\beta^{'}$ at that location. $K_s$ is sent to be verified by $\mathscr{H}_0$, which when $\mathscr{H}^{'}$ calls $K_s$ for a second time, under the given recursive property of $K_s$ which will eventually call itself, the z-check for value pair $(K_s, z)$ is recognized as both redundant and with a z value, stored by $\mathscr{H}^{'}$ in the data store, but because the associated value is z, the z-check ability tells us this process has already occurred, sends $K_s$ to $\mathscr{H}_1$, which self-verifies repeated z-check values. By construction, the only value $K_i$ which can provide this multiple z-check values where $K_i > c$ is $K_s$, so $\mathscr{H}_s$ now self-verifies the input $K_s$ as its own D.N., provides a value of ``$\emph{s}$" for satisfactory,
 and changes state to evaluate $K_s + 1$ to continue indefinitely as a \emph{Circle-free Turing Machine}. 
\\

Therefore,  given some Universal Turing Machine which can emulate the $\mathscr{H}_s$ Machine, $\beta^{'}$ is decidable over the set of given Description Numbers for all Standard Descriptions.
\end{proof}

\subsection{Consequences}

Solving for $\beta^{'}$ is RE-complete. If Turing's final result were correct, it would be impossible to evaluate a value of $u$ or $s$ for any given $\mathscr{H}$. However, we have constructed a machine which does exactly what Turing assumed was impossible. Doing so implies that ZFC is inconsistent. To thoroughly prove ZFC is inconsistent, we must not only show that there is a counter-example to the Halting problem, but exactly where the incorrect assumption appears in a logical fashion. We will find, in the following section, that there is an implied axiom of incompleteness in current logical implementations of ZFC where ZFC is open to contradiction.

\section{ZFC with Implied Axiom is Inconsistant}
\subsection{Incompleteness and Proof by Contradiction through Tautological Impredicatives}

G\"odel's second incompleteness theorem tells us that any formal system with the expressive power strong enough to represent the proof of its own consistency is either inconsistent or incomplete. This implies that in order to prove the consistency of a formal system such as ZFC, which has the expressive power to represent the proof of its own consistency through the formulation of Peano Postulates together with the Axiom of Substitution, must be incomplete if it is consistent. \cite{Godel}

We then assume that even though ZFC can express the proof of its own consistency, that it can not determine the proof of its own consistency, and as such, is incomplete, and we say such a proposition is independent of ZFC. However, while possibly true, this is not necessarily the case. G\"odel's second incompleteness theorem gives us a choice. It is possible that ZFC, and similarly expressive formal systems are in fact, inconsistent. \cite{Godel}

In order to prove ZFC is inconsistent, we must find two of its theorems which contradict each other. As stated earlier in this article, a proof of the existence of a Turing machine which can solve over $\beta^{'}$ implies ZFC is inconsistent. This is because many theorems in ZFC contradict such a finding.\footnote{The list of proofs of this kind is inexhaustible, however they include diagonalization arguments, such as not being able to calculate Kolmogorov Complexity, forcing techniques, Rice's Theorem, The Space-Hierarchy Theorem, et al.} The proofs of such theorems all have certain material similarities, and can be defined as a certain class of proof. First, the conclusion generalizes the non-existence of some statement or structure, $x$. Second, they are all proof by contradiction. Third, they all include an impredicative tautology. We can thus say that they are of the \emph{class of proof by contradiction through impredicative tautology}. We will informally describe, then formally define an impredicative tautology below.

\subsection{Impredicative Tautologies}

An impredicative statement is defined as a statement with self-referencing. Intuitively, it is easy to see how we risk creating a tautological consequence by using such a tactic to define one's terms: Self reference intrinsically eliminates the possibility for that variable to contradict itself, increasing the likelihood of tautology. 

Intuitively, we can see that Russell's paradox uses an impredicative definition that is itself also tautological. Russell's paradox depends on the definition of S as the set of all sets that do not contain themselves. This is impredicative, and it is tautological, because there seems to be this infinite self-referencing that goes on when we apply S to itself. The paradox arises as soon as we ask the question whether or not S is a member of itself or not. 

In Russell's paradox, the property of being a set that does not contain itself, is a property applied to a dependent portion of the impredicative statement, that is, the property when applied to sets in general, also applies to the specific set we are defining. This dependence of the property of the set on the set, is the distinguishing factor. The point here being that paradoxes of this type can be resolved by removing the tautological impredicative. One way of removing the tautological impredicative, is by defining a class as a collection of sets, and thus, the class of all sets that do not contain themselves, can not contain itself, because classes, by definition cannot contain classes, only sets.

But I believe this heuristic falls short. It is not just this dependence that creates a tautological impredicative statement, it is also the nature of the property itself. We could easily define the set of all sets which contain the letter A. Such a definition has a dependence on impredicative portion of the statement, yet does not seem to create create a tautology or an infinite regression or anything of that sort. That is, in order to create a tautology, ``A'' itself would have to be defined, not only in terms of sets, but in terms of the class of sets in question. As such, in order for impredicatives to be a problem for logic, the property itself must point back to the impredicative dependence in a self referential way. Let's call this \textit{impredicative pointing}.

So is there a more foundational way of removing tautological impredicatives?

\noindent \\
\textbf{Definition} Let \textit{impredicative dependence} be the condition of a statement S, whose property P depends on self-referencing. $\exists x | P(x) \leftrightarrow \{x\to P(x)\}$

\noindent \\
\textbf{Definition} Let \textit{impredicative pointing} be a condition of self-reference where a dependent property also references an impredicative dependence, i.e. the existence of S depends on S containing an impredicative dependence. $\exists x, S | P(S) \leftrightarrow P(x) \leftrightarrow \{x\to P(x)\} \to S$

\noindent \\
\textbf{Definition} Let a \textit{tautological impredicative} be a statement which satisfies impredicative pointing. $\exists x, S | P(S) \leftrightarrow P(x) \leftrightarrow \{x\to P(x)\} \to S$

\noindent \\
\textbf{Proposition} Tautological impredicatives are logical tautologies. $S\vDash \exists x, S | P(S) \leftrightarrow \{x\to P(x)\} \to S$\\

Consider the following truth table, $\forall x, \exists S$ such that:

\begin{center}
	\begin{tabular}{| l | l | l | l | l | l | p{4cm}|}
	\hline
	S & x & P(x) & x $\to$ P(x) & \{x $\to$ P(x)\} $\to$ S & P(S) & P(S) $\leftrightarrow$ \{x $\to$ P(x)\} $\to$ S\\ \hline
	T & T & T & T & T & T & T \\ \hline
	T & T & F & F & F & F & T \\ \hline
	T & F & T & F & F & F & T \\ \hline
	T & F & F & T & T & T & T \\ \hline
	F & T & T & T & F & F & T \\ \hline
	F & F & T & F & T & T & T \\ \hline
	F & F & F & T & F & F & T \\ \hline
	F & T & F & F & T & T & T \\ \hline
	\end{tabular}
\end{center}

As is clear, $P(S) \leftrightarrow \{x \to P(x)\} \to S$ is a tautology. $\square$
\\
\textbf{Proposition} Attempting to disprove existence with a proof by contradiction through tautological impredicative implies disjoint results which may or may not be the desired result.

Also note in the truth table above that when we assume $S$ is true, and by substitution we allow, $S = x$, $P(S)$ creates a contradiction by construction when $P(x)$ is false. This is because $P(S)$ is true when $S$ is true and $P(x)$ is false, but $x$ may or may not be false, setting up the contradiction $P(x)$ is false when $x$ is false. The logical consequence of this contradiction is either $\lnot S \oplus \exists x | P(x)$.

Therefore, in an attempt to prove $\lnot S$ using a proof by contradiction that relies on tautological impredicatives: \\
\\
$\forall x, \exists S | \lnot S \oplus P(x) \implies \lnot S \leftrightarrow \lnot P(x)$ \\
$P(x) \to S$ \\ $\square$\\
\noindent
\\
\textbf{Corollary} A contradiction may be the result of impredicative pointing, rather than the non-existence of a given $S$.\\
\\
When substitution is used to form a tautological impredicative in order to prove $\lnot S$ through proof by contradiction, one must also prove $\centernot \exists x | x \to P(x)$. Current practice is to assume this $x$ does not exist without proof. However, if it can be shown that $\exists x | x \to P(x) \implies P(x) \to S$, then this is sufficient to prove the existence of $S$, even if a contradiction still arises in the proof, as the contradiction does not arise if $\exists S$, the contradiction arises when $P(x) \land \lnot P(S) \leftrightarrow S=x$. Firmly placing contradiction in the hands of impredicative pointing, and not the initial assumption of existence for proof by contradiction. $\square$ \\

\noindent\textbf{Corollary} $S$ is false by contradiction only when we assume $\forall x| P(x)$ is false when $\exists S | P(S)$ is false. \\
\\
I want to be completely clear here: a contradiction arises from the use of substitution of $\forall x$ with $\exists S$ to form an impredicative tautology whether or not $x \to P(x) \lor \lnot \{x \to P(x)\}$. However, if $\lnot \{x \to P(x)\}$, then we can not be certain $S$ is false, because in this case, both $x$ and $S$ can be true when both $\lnot \{x \to P(x)\}$ and $\lnot P(S)$, retaining contradiction for proof. This means that the truth value for $P(x)$ is not determined through proof by contradiction; $P(x) $can be either true or false for any x. Therefore the conclusion that $S$ is false arises only when we must assume $\forall x| P(x)$ is false when $\exists S | P(S)$ is false. $\square$

Let's call such an assumption that $P(x)$ is false (or $\centernot\exists x$) when $\lnot P(S)$, while also assuming $S$ is true (for proof that $\lnot S$ by contradiction), a \textit{hidden assumption}.\footnote{As a corollary, S can also be the hidden assumption of a proof by contradiction through tautological impredicative when the proof openly assumes some property P(x) or x exists in order to disprove either P(x) or x.}

\textbf{Theorem} Turing's proof of the undecidability of the Halting Problem belongs to the class of proofs that are a proof by contradiction through impredicative tautology. \\

We may attempt to prove the undecidability of the Halting problem with the following simple version of a halting program: \\

\begin{verbatim}
def H():
    if halts(h):
        loop_forever()
\end{verbatim}

As we can see, we can use this program to prove the undecidability of the Halting problem through a proof by contradiction: If the subroutine $halts(h)$ halts, $h$ will loop forever, in which case $halts(h)$ is false. Let $h$ be the instructions for $H()$. If $H()$ halts, it will loop forever, which is a contradiction with $h$, which must halt to be satisfactory, in which case, $H(h)$ does not halt, which means it can not decide $h$, therefore $h$ is undecidable.

To show this proof by contradiction uses an impredicative tautology, we can designate the function $halts()$ as the property $P()$. We can let the instructions for $H()$, which contains $P()$, be $S$. We see that the proof defines $h=S$. However, because $h$ is not fixed in all cases, we may designate $h^{'}$ as some arbitrary definition for $H()$, $h^{'} \neq S$. 

\noindent\textbf{Proposition}. The Halting problem proof contains an impredicative dependence.
$\exists h | h \leftrightarrow \{h \to P(h)\}$ \\
We see this is true, because $h$ contains the instructions for $halts(x)$, which is $P(x)$ therefore $h \leftrightarrow \{h \to P(h)\}$.

\noindent\textbf{Proposition}.  The Halting problem proof contains impredicative pointing.\\
$\forall S, \exists h | P(S) \leftrightarrow P(h) \leftrightarrow \{h \to P(h)\} \to S$

We can see this is true, because first, $h \in S \implies P(S) \to P(S) \leftrightarrow P(h)$. That is $S$ contains $h$, so any property that applies to $S$, must also apply to $h$.

Second, $\exists S, h | \{h \leftrightarrow \{h \to P(h)\} \to S\} \to \{\{h \to P(h)\}\to S \implies S \to P(S)\}$ when $h = S$. That is, if $S$ only exists when there is impredicative dependence, then when $S$ exists in this manner, this implies when we substitute S with h, if $S$, then $P(S)$.

Third, This is enough to derive that the Halting problem contains impredicative pointing, since $\{\{S \to P(S)\}\land \{h \leftrightarrow \{h \to P(h)\} \to S\} \implies P(S) \leftrightarrow P(h) \leftrightarrow \{h \to P(h)\} \to S$ \\
\noindent\emph{Proof}. Finally, because the proof contains impredicative pointing, this means that the tautological impredicaive $H(h) \leftrightarrow \{x \to H(x)\} \to h$ is a logical consequence of the assumptions $S$ and $\lnot h^{'}$ through the formulation of the proof by contradiction. Thus, Turing's proof of the undecidability of the Halting problem belongs to the class of proofs that are a proof by contradiction through impredicative tautology. $\square$

\noindent\textbf{Corollary}. $\exists h^{'}$ serves as a counterexample to the proof of the undecidability of the Halting problem.

\noindent\textbf{Corollary}. If there exists a counterexample to the Halting problem, there exists a counterexample to the SPACE hierarchy theorem. It is well known that the SPACE hierarchy theorem reduces to the Halting problem. It immediately follows that if a counterexample to the Halting problem exists, then a counterexample to the SPACE hierarchy theorem exists.

\noindent\textbf{Proposition}. A counterexample to the SPACE hierarchy theorem exists. This is an immediate consequence of the above corollary and the proof of the existence of a counterexample to the Halting problem in section 1.  $\square$

\noindent\emph{Proof}. ZFC is inconsistent.

Since all current implementations of ZFC accept hidden assumptions in proofs by contradiction through tautological impredicative, then the following Axiom is implied by ZFC, even if not explicitly stated, by ZFC.

\noindent\textbf{Axiom}  $\forall x, \exists S | \{P(S) \leftrightarrow \{x \to P(x)\} \to S\} \to \forall S, \exists x | \{\lnot S \oplus P(x)\} \implies \lnot S$

In other words, ZFC implies a particular incompleteness where acceptance of $\centernot\exists x | P(x)$ in the circumstance of a proof of the undecidability of $S$ by contradiction through impredicative tautology.\footnote{The axiom of incompleteness described here is in widespread use by any and all mathematicians and computer scientists today.}

Accepting this axiom of incompleteness, as all logicians of note have since Post, Church and Turing, leads to contradiction when $\exists x | P(x)$. By the existence of the counterexample to Turing's proof in section 1, the direct implication is that $\exists x | P(x)$, which is in direct contradiction with the above axiom. It immediately follows that ZFC, with the implied axiom above, is inconsistent. $\square$

\subsubsection{A New Foundational Axiom}

We have demonstrated that substitution may lead to hidden assumptions in proof by contradiction. The need for a limit on how substitution is applied could help prevent such mistakes from occurring again. Perhaps we could just create a postulate or axiom which makes x a bounded variable after substitution. Such a postulate should allow some impredicative statements, all of which must avoid tautology, but through preventing certain re-substitutions on x, will prevent impredicative pointing, and thus prevent impredicative tautologies from forming. We can specify the limit of substitution over bounded x, not to substitution in general, but to statements from the free variable x. Thus, creating a much stronger foundation to our systems of logic and computability.
\\

\noindent\textbf{Axiom}. For any formal system $Q$, with free variables [x; y] and a substitution operation, subst(), z is bounded by subst() such that $\forall x,y,z | subst(x) = y  \land  subst(y) = z \to subst(z) \neq x.$\\

\section{Conclusion, P=NP}

In Section 1, we constructed a Supermachine which is able to solve for $\beta^{'}$. In section 2, we tackled the dangers of impredicative tautologies when applied to proofs by contradiction and proved that ZFC, by accepting hidden assumptions, is inconsistent. In this section, we will expand upon our findings to prove $P=NP$.

When proofs by contradiction utilizing an impredicative tautology are not allowed, we are no longer restricted by them. Furthermore, finding that the halting problem can be solved for all S.D. is in direct contradiction with the SPACE hierarchy theorem, because it is RE-complete. However, the SPACE hierarchy theorem is a proof by contradiction which utilizes an impredicative tautology to form a hidden assumption, which also reduces to the Halting problem which has a counterexample. Thus, the theorem is invalid by our findings. Similarly, as with other complexity separation arguments, we find that the entire complexity hierarchy collapses, and we now have a foundation in computer science where there is enough information to solve the P vs. NP problem. Without proof by contradiction with impredicative tautology, our reasons for not using an oracle to solve P vs. NP vanish, as the contradictions which formally prevented the use of such oracle no longer exist.\\

\noindent\emph{Lemma}. If $PSPACE=EXPSPACE$, $P=NP$. \\
If the $SPACE$ of a problem increases polynomially as with any $PSPACE$-complete problem, this is comparable to the $TIME$ of a problem increasing polynomially, such that given an oracle, $=_{Opoly}$,  which solves polynomial equivalence between SPACE and TIME, such that $PSPACE =_{Opoly} P$. Similarly, if the $SPACE$ of a problem increases exponentially as with any $EXPSPACE$-complete problem, this is comparable to NP which is at maximum, in exponential TIME, such that $EXPSPACE >=_{Opoly} NP$. If $PSPACE=EXPSPACE$, then $PSPACE>=_{Opoly} NP$. Since $PSPACE =_{Opoly} P$,  $P>=_{Opoly} NP$, which since P and NP are both in TIME, is the same as $P=NP$.

Solving for $\beta^{'}$ in Section 1 is RE-complete, and because the SPACE hierarchy theorem relies fully on the now defunct method of proof by contradiction utilizing an impredicative tautology, its results \textbf{\emph{must}} be discarded. 

And as such, with the Halting problem being RE-complete, and since we may solve for arbitrary $\beta^{'}$ using the Supermachine configuration, we may now conclude:
	\begin{proof}
		Since by definition, $PSPACE\subseteq RE$, and  
\\
		since any given Recursively Enumerable set is contained in PSPACE,
\\
		and $\beta^{'}$ solves for all Recursively Enumerable sets in PSPACE,
\\
		and since we can no longer accept the SPACE hierarchy theorem, 
\\
		$RE\subseteq PSPACE$ ...
\\
		$RE = PSPACE$,
\\
		such that $EXPSPACE\subseteq RE$
\\
		and $RE = PSPACE$, implies
\\
		$PSPACE=EXPSPACE$, proves through the above Lemma ...
\\
		$P = NP$
	\end{proof}

\begin{flushright}
Mark Inman, Ph.D.\\
Saipan, Northern Mariana Islands, USA
\end{flushright}

\end{document}